\documentclass[12pt]{article}
\usepackage{amsmath}
\usepackage{amssymb}
\usepackage{graphicx}
\usepackage{url}
\usepackage[small]{caption}
\let\oldFootnote\footnote
\newcommand\nextToken\relax

\renewcommand\footnote[1]{%
    \oldFootnote{#1}\futurelet\nextToken\isFootnote}

\newcommand\isFootnote{%
    \ifx\footnote\nextToken\textsuperscript{,}\fi}

\begin{document}
\baselineskip 0.6cm

\def\simgt{\mathrel{\lower2.5pt\vbox{\lineskip=0pt\baselineskip=0pt
           \hbox{$>$}\hbox{$\sim$}}}}
\def\simlt{\mathrel{\lower2.5pt\vbox{\lineskip=0pt\baselineskip=0pt
           \hbox{$<$}\hbox{$\sim$}}}}
\def\simprop{\mathrel{\lower3.0pt\vbox{\lineskip=1.0pt\baselineskip=0pt
             \hbox{$\propto$}\hbox{$\sim$}}}}
\def\bra#1{\left< #1 \right|}
\def\ket#1{\left| #1 \right>}
\def\inner#1#2{\left< #1 | #2 \right>}

\begin{titlepage}

\begin{flushright}
MIT-CTP-4425 \\
UCB-PTH-12/19
\end{flushright}

\vskip 1.6cm

\begin{center}
{\Large \bf A Note on (No) Firewalls: The Entropy Argument}

\vskip 0.7cm

{\large Yasunori Nomura and Jaime Varela}

\vskip 0.4cm

{\it Center for Theoretical Physics, Laboratory for Nuclear Science, 
     and Department of Physics, \\
     Massachusetts Institute of Technology, Cambridge, MA 02139, USA} \\

\vskip 0.2cm

{\it Berkeley Center for Theoretical Physics, Department of Physics, \\
     and Theoretical Physics Group, Lawrence Berkeley National Laboratory, \\
     University of California, Berkeley, CA 94720, USA}

\vskip 0.8cm

\abstract{An argument for firewalls based on entropy relations is refuted.}

\end{center}
\end{titlepage}

Recently, Almheiri, Marolf, Polchinski, and Sully (AMPS) have posed an 
interesting paradox~\cite{Almheiri:2012rt}:\ the equivalence principle 
for an infalling observer is incompatible with the hypothesis that the 
formation and evaporation of a black hole, as viewed from a distant 
observer, is described by unitary quantum mechanics and that physics 
outside the stretched horizon is well approximated by a semi-classical 
theory (the complementarity hypothesis~\cite{Susskind:1993if}).  If 
true, this would have profound implications for physics of spacetime 
and gravity.  In particular, AMPS advocate that the most conservative 
resolution of the paradox is that for an old (but still macroscopic) 
black hole, the infalling observer hits a ``firewall'' of high energy 
quanta at the horizon---a drastic deviation from the prediction of 
general relativity.

AMPS present two arguments for firewalls:\ one based on a measurement 
of early Hawking radiation by an observer falling into an old 
black hole and the other based on properties of entropies associated 
with various subsystems of an old black hole.  In previous papers, 
we, together with Sean Weinberg, have refuted the first AMPS 
argument~\cite{Nomura:2012sw,Nomura:2012cx}.  (For other work 
on firewalls, see, e.g.,~\cite{Bousso:2012as}.)  A key observation 
is that a {\it full} quantum state, to which the unitarity argument 
applies, is in general a superposition of different classical worlds; 
on the other hand, general relativity (or the equivalence principle) 
applies to {\it each} of these classical worlds, not necessarily to 
the whole quantum state.  In this short note, we show that a similar 
observation also allows us to avoid the second AMPS argument, i.e.\ 
the argument based on entropies.

Let us briefly recall how the first AMPS argument was addressed in 
Refs.~\cite{Nomura:2012sw,Nomura:2012cx}.  The argument is concerned 
about a state of an old black hole (after the Page time) that has formed 
from collapse of some pure state:
\begin{equation}
  \ket{\Psi} = \sum_i c_i \ket{\psi_i} \otimes \ket{i},
\label{eq:state-1}
\end{equation}
where $\ket{\psi_i} \in {\cal H}_{\rm rad}$ and $\ket{i} \in 
{\cal H}_{\rm horizon}$ represent degrees of freedom associated with 
early Hawking radiation and the horizon region, respectively.  Now, since 
${\rm dim}\,{\cal H}_{\rm rad} \gg {\rm dim}\,{\cal H}_{\rm horizon}$, 
one can construct an operator $P_i$ that acts only on ${\cal H}_{\rm rad}$ 
(not on ${\cal H}_{\rm horizon}$) but selects a term in Eq.~(\ref{eq:state-1}) 
corresponding to {\it any} state $\ket{i} \in {\cal H}_{\rm horizon}$: 
$P_i \ket{\Psi} \propto \ket{\psi_i} \otimes \ket{i}$.  Suppose we 
choose $P_i$ so that $\ket{i}$ is an eigenstate of the number operator, 
$b^\dagger b$, of a Hawking radiation mode that will escape from the 
horizon region.  The state $\ket{i}$ then cannot be a vacuum for the 
infalling modes $a_\omega$, which are related to $b$ by $b = \int_0^\infty\! 
d\omega ( B(\omega) a_\omega + C(\omega) a_\omega^\dagger)$ with some 
functions $B(\omega)$ and $C(\omega)$.  AMPS argue that the fact that 
one {\it can} choose such $P_i$ implies that the infalling observer 
{\it must} encounter high energy modes, i.e.\ the firewall, because 
he/she can (in principle) measure early Hawking radiation to select 
the particular term $\ket{\psi_i} \otimes \ket{i}$ in $\ket{\Psi}$.

This argument, however, misses the fact that the existence of projection 
operator $P_i$ does {\it not} mean that a measurement performed by 
a {\it classical} observer, which general relativity is supposed to 
describe, picks up the corresponding state $\ket{i}$.  In fact, quite 
generally, the state $\ket{\psi_i}$ that is entangled with a $b^\dagger b$ 
eigenstate $\ket{i}$ is a superposition of states $| \hat{\psi}_a \rangle$ 
having well-defined classical configurations of Hawking radiation quanta:
\begin{equation}
  \ket{\psi_i} = \sum_a d_{ia} |\hat{\psi}_a \rangle,
\label{eq:psi_i}
\end{equation}
with $d_{ia} \not\approx \delta_{ia}$~\cite{Nomura:2012sw}.  An important 
point is that the coefficients $d_{ia}$ are determined dynamically by 
the form of Hamiltonian, especially its local nature---it is not something 
we can choose arbitrarily, e.g.\ as $d_{ia} = \delta_{ia}$, independent of 
the dynamics.  Substituting Eq.~(\ref{eq:psi_i}) into Eq.~(\ref{eq:state-1}), 
we obtain
\begin{equation}
  \ket{\Psi} = \sum_a |\hat{\psi}_a \rangle \otimes 
    \left( \sum_i d_{ia} c_i \ket{i} \right) 
  \equiv \sum_a |\hat{\psi}_a \rangle \otimes |\hat{a} \rangle,
\label{eq:state-2}
\end{equation}
implying that $\ket{\Psi}$ is a superposition of terms having well-defined 
configurations of Hawking quanta.  Now, a classical world can be 
defined as a basis state in which the information is amplified; see, 
e.g.,~\cite{q-Darwinism,Nomura:2011rb}.  In Ref.~\cite{Nomura:2012sw}, 
it was considered that the natural basis in the present context, i.e.\ 
in addressing AMPS's first argument, is given by interactions between 
Hawking quanta and the classical measuring devise, which is spanned by 
the $|\hat{\psi}_a \rangle$'s.  General relativity says that the horizon 
state in each classical world must be approximately a vacuum state for 
the infalling modes, and this is not a contradiction since $|\hat{a} 
\rangle$ can be far from an eigenstate of $b^\dagger b$: $|\hat{a} 
\rangle \not\approx \ket{i}$.%
\footnote{If we prepare a carefully-crafted quantum device that will 
 be entangled with one of the $\ket{i}$'s and send, e.g., a particle 
 toward the horizon, then that particle may see a firewall.  Such a 
 device, however, needs to be an exponentially fine-tuned superposition 
 state of different classical configurations, and can be ignored under 
 realistic circumstances~\cite{Nomura:2012cx}.}

We now turn to the main theme of the present note.  We suggest that 
the resolution of the firewall paradox may lie in the emergence of 
classical worlds in a full quantum state, and this can be determined 
by the {\it internal dynamics of the horizon} (when the system is viewed 
from outside).  (The role played by interactions of a device with early 
Hawking quanta can be minor.)  In particular, the fallen object is 
represented differently {\it at the microscopic level} in each of these 
classical worlds, although they all correspond to the object falling 
in the same infalling vacuum when described in general relativity. 
We will see that this can address AMPS's second argument based on 
entanglements (and their first argument as well).

AMPS's second argument goes as follows.  Consider three subsystems of 
an old black hole $A$, $B$, and $C$.  In an infalling frame, take
\begin{align}
  A:&\,\,\, \mbox{early/distant Hawking modes,}
\label{eq:A}\\
  B:&\,\,\, \mbox{outgoing modes localized near outside of a (small) 
    patch of the horizon,}
\label{eq:B}\\
  C:&\,\,\, \mbox{modes inside the horizon that are Hawking partners 
    of $B$.}
\label{eq:C}
\end{align}
In a distant frame, the {\it interpretation} of $C$ (but not of $A$ 
or $B$) changes:
\begin{align}
  C:&\,\,\, \mbox{a subsystem of the degrees of freedom composing 
    the stretched horizon,}
\label{eq:C-distant}
\end{align}
although it still represents the {\it same} degrees of freedom as in 
Eq.~(\ref{eq:C}) (complementarity).  Now, unitarity says that for an old 
black hole the entropy of the distant modes decreases, implying
\begin{equation}
  S_{AB} < S_A,
\label{eq:ent-rel-1}
\end{equation}
where $S_X$ represents the von~Neumann entropy of system $X$.  On 
the other hand, the equivalence principle applied to a freely falling 
observer says
\begin{equation}
  S_{BC} = 0,
\label{eq:ent-rel-2}
\end{equation}
implying $S_{ABC} = S_A$.  These two relations contradict strong 
subadditivity of entropy
\begin{equation}
  S_{AB} + S_{BC} \geq S_B + S_{ABC},
\label{eq:ent-rel-3}
\end{equation}
since they lead to $S_B < 0$ if both are true.  This implies that one 
must abandon either unitarity of a black hole formation/evaporation 
process (with physics outside the stretched horizon well described by 
a semi-classical theory) or the equivalence principle.

What can be wrong with this argument?  Again, the key observation is 
that unitarity is a statement about an entire quantum state while the 
equivalence principle is a statement about a classical world---a 
component/branch of the entire quantum state.  Let us assume that 
both these statements are correct.  Then, for the entire quantum state, 
Eq.~(\ref{eq:ent-rel-1}) must apply while Eq.~(\ref{eq:ent-rel-2}) 
need not (and in fact cannot) be satisfied.  On the other hand, if 
we focus only on a single component of the state corresponding to 
a classical world, then the relation as in Eq.~(\ref{eq:ent-rel-2}) 
must be satisfied (where the entropy should be derived only from that 
particular component of the state, which we will denote as $\tilde{S}_{BC}$ 
from now on), while the bound as in Eq.~(\ref{eq:ent-rel-1}), i.e.\ 
$\tilde{S}_{AB} < \tilde{S}_A$, will not be true.  This is consistent 
because the information is (almost) always lost for a classical observer 
in a quantum mechanical system.

To illustrate this point more explicitly, let us consider the quantum 
state of an old black hole, which we write in the form
\begin{equation}
  \ket{\Psi} = \sum_{i,j,k,l} c_{ijkl} \ket{A_i}\ket{B_j}\ket{C_k}\ket{D_l},
\label{eq:BH-state}
\end{equation}
where $\ket{B_j}$ and $\ket{C_k}$ represent states for $B$ and $C$ in 
Eqs.~(\ref{eq:B},~\ref{eq:C},~\ref{eq:C-distant}).  (Here and below, 
we omit the direct-product symbol.)  $\ket{D_l}$ are states for subsystem 
$D$ which represents all the internal (or stretched horizon) degrees 
of freedom other than $C$, and $\ket{A_i}$ are states for $A$, in which 
we now include all the outside degrees of freedom other than $B$. 
The state $\ket{\Psi}$ comprises our entire quantum state.%
\footnote{To be precise, $\ket{\Psi}$ may not be the {\it complete} state 
 for an old black hole, which in general is a superposition of various 
 $\ket{\Psi}$'s corresponding to black holes in different locations 
 and with different spins~\cite{Nomura:2012cx}.  Therefore, the complete 
 information about the initial state may not be reproduced from a single 
 $\ket{\Psi}$ alone.  This aspect, however, is irrelevant for our 
 argument below, since the entropy for distant radiation decreases 
 in $\ket{\Psi}$ (not only in the complete state), i.e.\ $S_{AB} 
 < S_A$ in $\ket{\Psi}$.}

Unitarity of the evolution of $\ket{\Psi}$ implies that 
Eq.~(\ref{eq:ent-rel-1}) must apply.  The strong subadditivity 
relation in Eq.~(\ref{eq:ent-rel-3}) then says that Eq.~(\ref{eq:ent-rel-2}) 
cannot be true.  In fact, to satisfy the relation, $S_{BC}$ must be 
of order $S_B$, so the $BC$ system is far from maximally entangled. 
Namely, we have
\begin{equation}
  S_{AB} < S_A,
\qquad
  S_{BC} \not\approx 0.
\label{eq:ent-Psi}
\end{equation}
Does this mean that general relativity is incorrect, i.e.\ a freely 
falling {\it classical} observer finds a drastic violation of the 
equivalence principle at the horizon?

The answer is no.  As discussed before, emergence of classical worlds 
in a quantum mechanical system is controlled by the dynamics, and 
a quantum state is in general a superposition of these classical worlds. 
In particular, the dynamics selects a set of natural basis states in 
which the information is amplified, and with which any classical objects 
would be entangled.  (For a sufficiently large system with a local 
Hamiltonian, the basis states are those having well-defined configurations 
in classical phase space, with spreads dictated by the uncertainty 
principle.)  Let us now consider a state representing one of these 
classical worlds
\begin{equation}
  |\tilde{\Psi} \rangle = z \sideset{}{'}\sum_{i,j,k,l} 
    c_{ijkl} \ket{A_i}\ket{B_j}\ket{C_k}\ket{D_l},
\label{eq:BH-state-classical}
\end{equation}
where the sum runs only over a subset of the $A$ through $D$ states so 
that $ |\tilde{\Psi} \rangle$ corresponds to a (decohered) classical world 
in $\ket{\Psi}$.    (The sum is denoted with prime to emphasize this 
point, and $z$ is the normalization constant.)  We can define von~Neumann 
entropies for subsystems $A$, $B$, etc.\ {\it of the state $|\tilde{\Psi} 
\rangle$} (not of $\ket{\Psi}$).  We call such entropies {\it branch 
world entropies} and denote them with the tilde: $\tilde{S}_X$ for 
a subsystem $X$ of $|\tilde{\Psi} \rangle$.

The validity of general relativity requires the $BC$ system in 
$|\tilde{\Psi} \rangle$ to be maximally entangled, and applying 
the strong subadditivity relation to $|\tilde{\Psi} \rangle$ then leads 
to the conclusion that the entropy of the combined $AB$ system cannot 
be lower than that of $A$:
\begin{equation}
  \tilde{S}_{BC} = 0,
\qquad
  \tilde{S}_{AB} \nless \tilde{S}_A.
\label{eq:ent-Psi-tilde}
\end{equation}
In fact, with $\tilde{S}_{BC} = 0$, the strong subadditivity relation 
yields a stronger condition $\tilde{S}_{AB} \geq \tilde{S}_A + \tilde{S}_B$. 
Together with another basic inequality of entropy $\tilde{S}_{AB} \leq 
\tilde{S}_A + \tilde{S}_B$, this leads to
\begin{equation}
  \tilde{S}_{AB} = \tilde{S}_A + \tilde{S}_B,
\label{eq:separable}
\end{equation}
i.e.\ two subsystems $A$ and $B$ are (almost) separable in a classical 
world in which $\tilde{S}_{BC} \approx 0$ is valid.  This implies that 
a classical observer in $|\tilde{\Psi} \rangle$ cannot see any entanglement 
between $A$ and $B$.

We now come to our main point:\ relations in Eq.~(\ref{eq:ent-Psi}) are 
{\it not} incompatible with those in Eq.~(\ref{eq:ent-Psi-tilde})---they 
are relations on two different quantities:\ entropies for the entire 
state and branch world entropies.  The two are compatible because for 
an old black hole the coefficients $c_{ijkl}$ in Eq.~(\ref{eq:BH-state}) 
can have significant support spanning different classical worlds 
(otherwise $\ket{\Psi} \approx |\tilde{\Psi} \rangle$) and because 
the infalling ``vacuum'' state in the $BC$ region need not be unique 
(otherwise, the $BC$ state could be factored from $\ket{\Psi}$, making 
$S_{AB} < S_A$ impossible to satisfy).  Here, different classical worlds 
mean different microstates for the $BC$ states that are described as 
the same semi-classical spacetime in general relativity.

In the true Minkowski space, the vacuum state is believed to be unique. 
The near horizon region, however, is only Minkowski~vacuum-like (i.e.\ 
the equivalence principle requires only a small region compared with 
the black hole to be Minkowski~vacuum-like), and there can be many such 
states because of microscopic degrees of freedom of the black hole. 
In particular, the vacuum state for each classical branch can differ 
in which subsystem of the black hole (i.e.\ $C + D$) is identified as 
$C$ (i.e.\ the partner modes of $B$) and/or how the modes in $B$ are 
entangled with those in $C$.  (General relativity, however, describes 
all these states as the same infalling vacuum.)  Branches having different 
entanglement structures between $B$ and the black hole degrees of freedom 
correspond to different decohered classical worlds.

Because the dimension of Hilbert space for the $CD$ system (black hole) 
is $e^{{\cal A}/4 l_P^2}$, where ${\cal A}$ and $l_P$ are the area of 
the black hole and the Planck length respectively, the number of different 
classical worlds can be as large as $e^{{\cal A}/4 l_P^2}$, enough to fully 
recover unitarity.  The basis for the classical states is selected by the 
internal dynamics of the horizon, which we assume to be maximally-entangled 
black hole and exterior near-horizon mode states corresponding to the 
infalling vacuum.  Overlaps among these states are extremely small, so 
they are regarded as different decohered classical worlds.  An infalling 
classical (macroscopic) object will be entangled with states in this basis. 
In particular, the object entering the horizon is represented by different 
states of the black hole in various branches, specifically as a small 
fluctuation around each of the $e^{{\cal A}/4 l_P^2}$ different vacuum 
states.

For ${\rm dim}\,{\cal H}_A \simgt {\rm dim}\,{\cal H}_{BC}$, one can 
easily see that $\ket{\Psi}$ can satisfy Eq.~(\ref{eq:ent-Psi}) while 
keeping Eq.~(\ref{eq:ent-Psi-tilde}) for each classical state $|\tilde{\Psi} 
\rangle$.  Of course, this does not prove what we have postulated above:\ 
(i) unitarity, (ii) the equivalence principle, and (iii) the semi-classical 
nature of physics outside the stretched horizon---these are still 
assumptions.  It, however, does show that the argument by AMPS is 
flawed, and that (i), (ii), and (iii) can all simultaneously be true.

Once again, an important point is to realize that states corresponding 
to well-defined classical worlds are very special in quantum mechanics 
(most of the states in general Hilbert space are superpositions of 
different classical worlds) and that general relativity is a theory 
describing a classical world (whose emergence is controlled by the dynamics 
of a system), as has been emphasized by the authors both in black hole 
physics~\cite{Nomura:2012sw,Nomura:2012cx} and cosmology (especially 
the eternally inflating multiverse)~\cite{Nomura:2011dt,Nomura:2011rb}. 
By carefully considering this point, we conclude that complementarity 
is a consistent hypothesis.  A realization of it in which the 
intrinsically quantum mechanical nature is manifest has been discussed 
in Ref.~\cite{Nomura:2011rb}, where complementarity is interpreted as 
(a part of) unitary reference-frame change transformations acting on 
the covariant Hilbert space.  For detailed discussions on this proposal 
in the context of black holes, see~\cite{Nomura:2012cx}.

\section*{Acknowledgments}

We thank Jared Kaplan for useful discussions.  This work 
was supported in part by the Director, Office of Science, Office 
of High Energy and Nuclear Physics, of the US Department of Energy 
under Contracts DE-FG02-05ER41360 and DE-AC02-05CH11231, by the 
National Science Foundation under grants PHY-0855653 and DGE-1106400, 
and by the Simons Foundation grant 230224.

\end{document}